\documentclass[aps,prd,groupedaddress,graphicx,nofootinbib]{revtex4}
\usepackage{amsmath,amssymb,graphics,graphicx,color,epsf}
\usepackage{subfigure}
\usepackage[scanall]{psfrag}  
\usepackage{tikz}

\begin{document}

\title{Chaotic inflation on the brane and the Swampland Criteria}

\author{Chia-Min Lin$^{1}$}
\author{Kin-Wang Ng$^{2,3}$}
\author{Kingman Cheung$^{4,5,6}$}
\affiliation{$^1$Fundamental Education Center, National Chin-Yi University of Technology, Taichung 41170, Taiwan}
\affiliation{$^2$Institute of Physics, Academia Sinica, Taipei 11529, Taiwan}
\affiliation{$^3$Institute of Astronomy and Astrophysics, Academia Sinica, Taipei 11529, Taiwan}
\affiliation{$^4$Depaertment of Physics, National Tsing Hua University, Hsinchu 300, Taiwan}
\affiliation{$^5$Physics Division, National Center for Theoretical Sciences, Hsinchu 300, Taiwan}
\affiliation{$^6$Division of Quantum Phases and Devices, School of Physics, Konkuk University, Seoul 143-701, Republic of Korea}


\date{Draft \today}

\begin{abstract}
In this paper, we show that single-field chaotic inflation on the brane with the potential $V=a \phi^p$ is compatible with the Swampland criteria. The spectral index and the running spectral index are within experimental bounds for $0<p \leq 2$. The tensor to scalar ratio is within observational bounds if $p \lesssim \mathcal{O}(1)$.
\end{abstract}
\maketitle
\large
\baselineskip 18pt
\section{Introduction}

If we have a UV completed theory, it should be able to give some insight to our low-energy effective theory. Even if we do not have a UV completed theory yet, if the theory is believed to have some general properties, it could shed some light to our model building. Recenlty it was proposed that if an effecitve field theory can be embedded consistently in quantum gravity, it has to satisfy two criteria \cite{Ooguri:2006in,Obied:2018sgi}. These criteria provide such strong constraints, that they are incompatible to many inflation models \cite{Kinney:2018nny, Achucarro:2018vey, Garg:2018reu}. 
These two criteria are as follows:
\begin{itemize}
\item \textbf{The Swampland distance conjecture:}
\begin{equation}
\frac{\Delta \phi}{M_P} < \mathcal{O}(1),
\end{equation}
which states that scalar field excursion in reduced Planck units in field space are bounded from above \cite{Ooguri:2016pdq}.
\item \textbf{The Swampland de Sitter Conjecture:}
\begin{equation}
M_P \frac{|V^\prime|}{V}>c \sim \mathcal{O}(1),
\end{equation}
which states that the slope of the scalar field potential satisfies a lower bound whenever $V>0$ \cite{Agrawal:2018own} and the bounds are of $\mathcal{O}(1)$\footnote{The swampland conjectures are basically applicable to any dimensions except for low ones where the gravity is not dynamical. The coefficient  $c$  in the conjecture can depend on the dimension $d$, though the value of $c \sim O(1)$ for $d=4$ or $d=5$ \cite{Obied:2018sgi}. Nevertheless, only the $d=4$, the current version, is interesting phenomenologically and can be applied directly to inflationary models.}.
\end{itemize}

The number of e-folds of conventional single field slow roll inflation is given by
\begin{equation}
N= \frac{1}{M_P^2}\int \frac{V}{V^\prime}d\phi \simeq \frac{\frac{\Delta \phi}{M_P}}{M_P \frac{V^\prime}{V}},
\end{equation}
where the approximation is obtained by assuming $V^\prime/V$ is independant of $\phi$.
Therefore for single field inflation the first criterion divided by the second one roughly gives the number of e-folds during inflation. This suggests that it may not be possible to have a large enough number of e-folds. In addition,
if the primordial density perturbation comes from inflaton fluctuation, the second constaint seems to be incompatible with current observation of tensor to scalar ratio.

One possibility to evade those criteria in single-field inflation is to consider a curvaton-like mechanism \cite{Kehagias:2018uem}.
Another possible way is given in \cite{Brahma:2018hrd}, where it is pointed out that quintessential brane inflation is compatible with swampland criteria, however, the unacceptably high tensor to scalar ratio is predicted unless the initial state have a non-Bunch-Davies component. In addition, it is shown in \cite{Das:2018hqy} that warm inflation or k-inflation may also be compatible with the criteria.

In this paper, we consider chaotic inflation on the brane and show that it can satisfy both swampland criteria and the tensor to scalar ratio constraints.

\large
\baselineskip 18pt

\section{Inflation on the brane}
\label{sec2}
In a braneworld scenario where our four-dimensional world is a 3-brane embedded in a higher-dimensional bulk, the Friedmann equation can be modified as \cite{Cline:1999ts,Csaki:1999jh,Binetruy:1999ut,Binetruy:1999hy,Freese:2002sq,Freese:2002gv,Maartens:1999hf}
\begin{equation}
H^2=\frac{1}{3M_P}\rho \left[ 1+\frac{\rho}{2 \Lambda} \right],
\end{equation} 
where $\Lambda$ provides a relation between the four-dimensional Planck scale $M_4$ and five-dimensional Planck scale $M_5$ through
\begin{equation}
M_4=\sqrt{\frac{3}{4\pi}} \left( \frac{M_5^2}{\sqrt{\Lambda}} \right) M_5,
\end{equation}
where $M_P=M_4/\sqrt{8\pi} \simeq 2.4 \times 10^{18}$ GeV is the reduced Planck scale. We will set $M_P=1$ in the following. The nucleosynthesis limit implies that $\Lambda \gtrsim (1 \mbox{ MeV})^4 \sim (10^{-21})^4$. A more stringent constraint, $M_5 \gtrsim 10^5$ TeV, can be obtained by requiring the theory to reduce to Newtonian gravity on scales larger than 1 mm, this corresponds to $\Lambda \gtrsim 5.0 \times 10^{-53}$.

For inflation on the brane, the slow-roll parameters are modified into \cite{Maartens:1999hf}
\begin{eqnarray}
\epsilon &\equiv& \frac{1}{2} \left( \frac{V^\prime}{V} \right)^2 \frac{1}{\left( 1+\frac{V}{2\Lambda} \right)^2}\left( 1+\frac{V}{\Lambda} \right), \label{epsilon} \\
\eta &\equiv& \left( \frac{V^{\prime\prime}}{V} \right)\left( \frac{1}{1+\frac{V}{2 \Lambda}} \right). \label{eta}
\end{eqnarray}
The number of e-folds is 
\begin{equation}
N=\int^{\phi_i}_{\phi_e} \left( \frac{V}{V^\prime} \right) \left( 1+\frac{V}{2 \Lambda} \right) d\phi.
\label{efolds}
\end{equation}
The spectrum is 
\begin{equation}
P_R=\frac{1}{12\pi^2}\frac{V^3}{V^{\prime 2}} \left( 1+\frac{V}{2 \Lambda} \right)^3. 
\label{spectrum}
\end{equation} 
The spectral index is
\begin{equation}
n_s=1+2 \eta -6 \epsilon.
\label{index}
\end{equation}
The running spectral index is given by
\begin{equation}
\alpha = \frac{d n_s}{d \ln k}= -\frac{d n_s}{d N}.
\end{equation}

\section{Chaotic Inflation on the brane}
\begin{figure}[t]
\centering
\includegraphics[width=0.7\columnwidth]{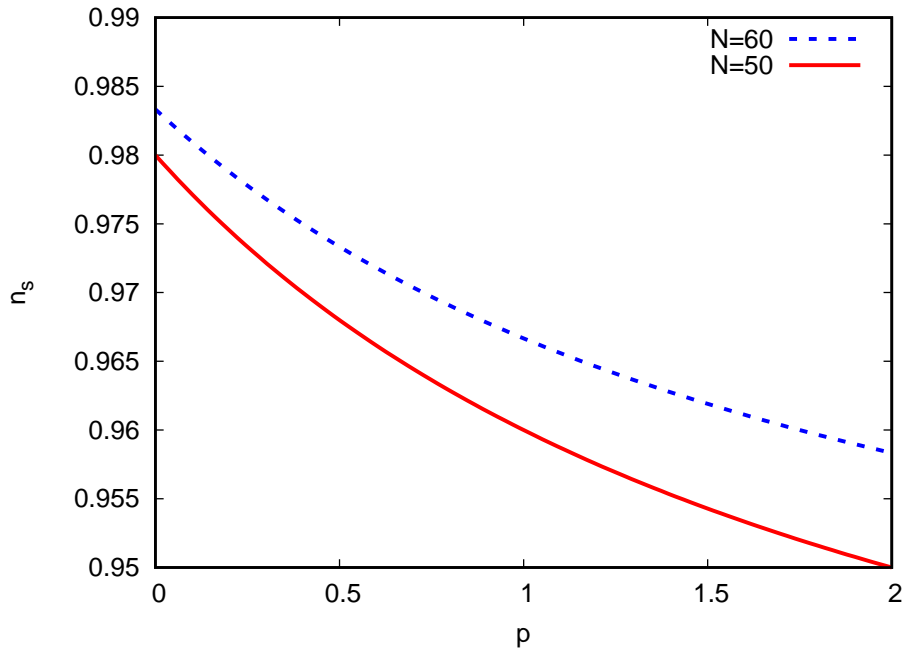}
 \caption{The spectral index $n_s$ as a function of $p$. Both number of e-folds $N=50$ and $60$ are shown.}
\label{fig7}
\end{figure}

\begin{figure}[t]
\centering
\includegraphics[width=0.7\columnwidth]{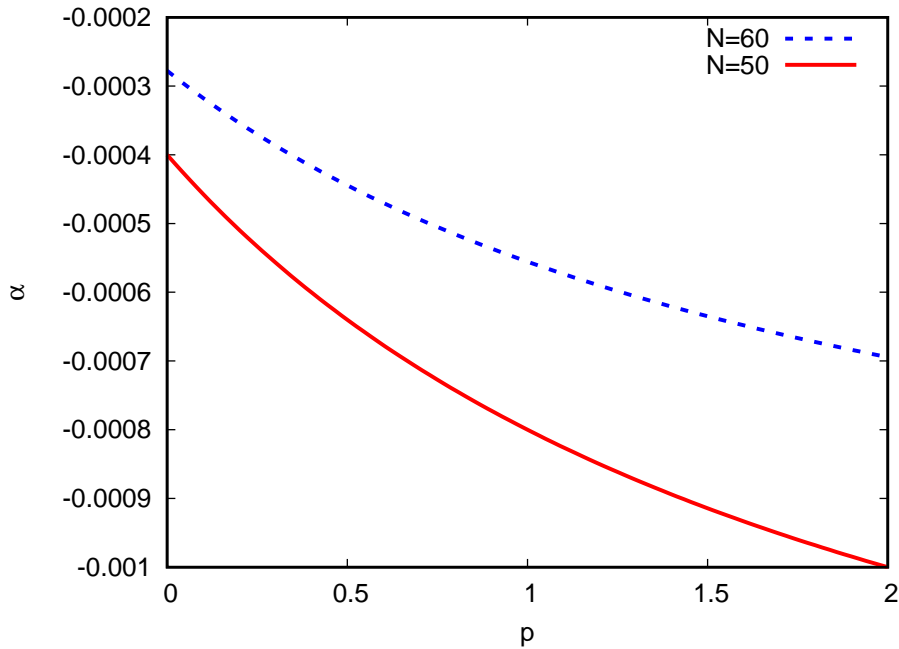}
 \caption{The running spectral index $\alpha$ as a function of $p$. Both number of e-folds $N=50$ and $60$ are shown.}
\label{running}
\end{figure}

\begin{figure}[t]
\centering
\includegraphics[width=0.7\columnwidth]{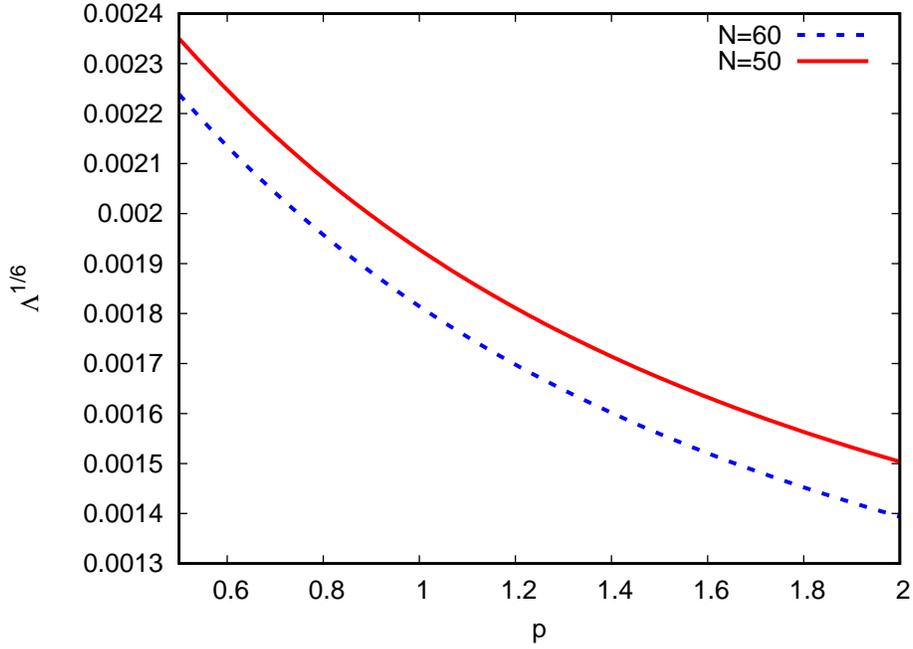}
 \caption{The upper bound of $\Lambda^{1/6}$ as a function of $p$ when $0.5 \leq p \leq 2$ from the condition $V/\Lambda > 100$. Both number of e-folds $N=50$ and $60$ are shown.}
\label{fig5}
\end{figure}

\begin{figure}[t]
\centering
\includegraphics[width=0.7\columnwidth]{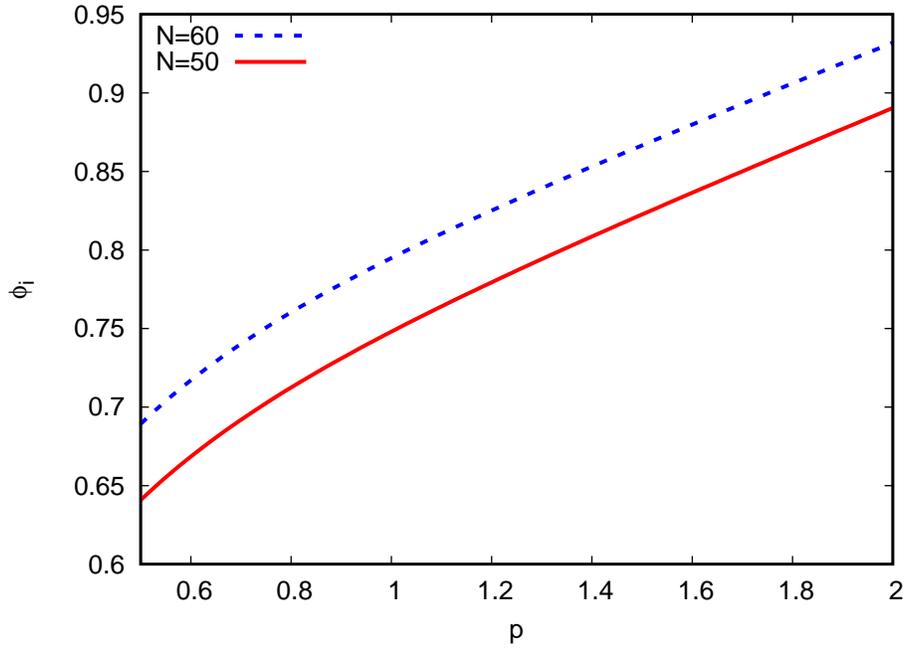}
 \caption{The upper bound of $\phi_i$ as a function of $p$ when $0.5 \leq p  \leq 2$ from the condition $V/\Lambda > 100$. Both number of e-folds $N=50$ and $60$ are shown.}
\label{fig1}
\end{figure}

\begin{figure}[t]
\centering
\includegraphics[width=0.7\columnwidth]{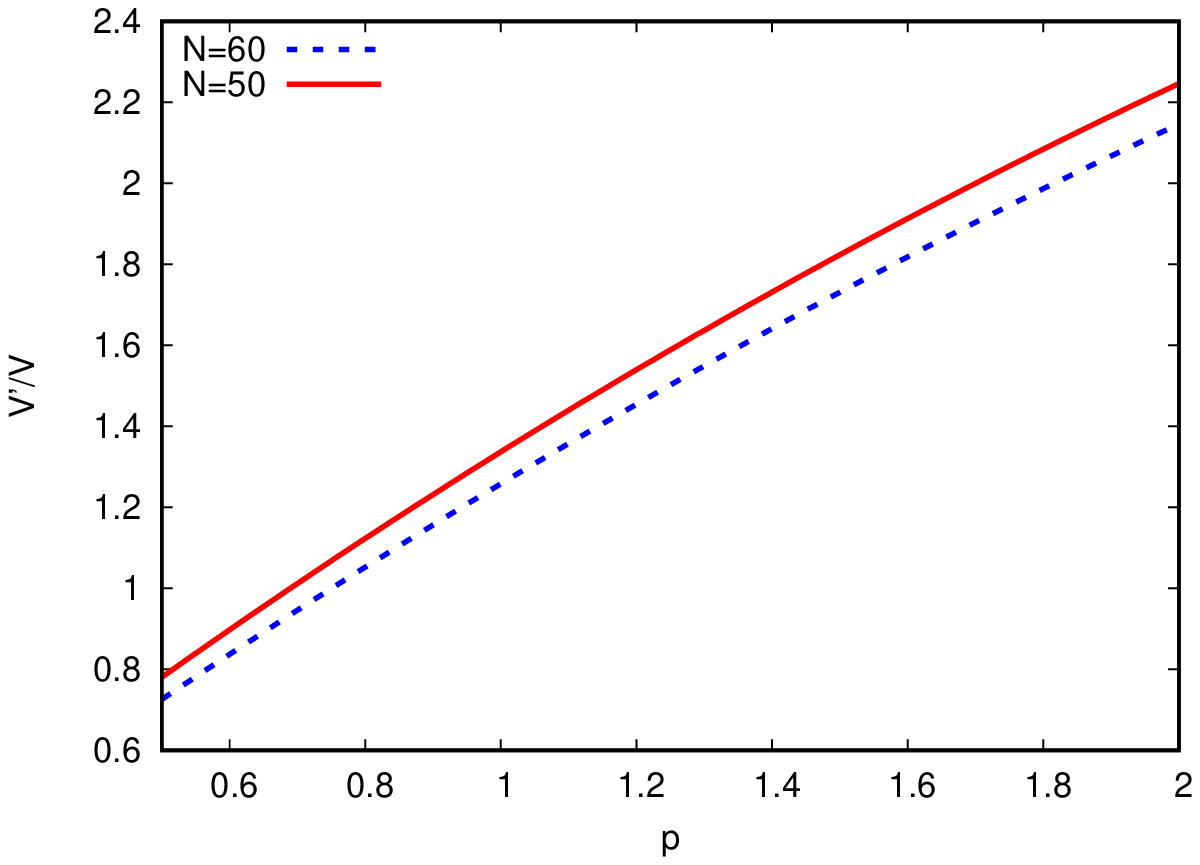}
 \caption{The lower bound of $V^\prime/V$ as a function of $p$ when $0.5 \leq p \leq 2$ from the condition $V/\Lambda > 100$. Both number of e-folds $N=50$ and $60$ are shown.}
\label{fig4}
\end{figure}

Let us consider chaotic inflation on the brane with a potential of the form\footnote{Monomial potentials with fractional powers have been considered in sring theory \cite{Silverstein:2008sg, McAllister:2008hb, Kaloper:2008fb, McAllister:2014mpa} where models with $p=2/5,2/3,1,4/3$ are exhibited. We will assume $0 <p \leq 2$ in this work.} 
\begin{equation}
V=a \phi^p.
\end{equation}
From the potential we can obtain
\begin{eqnarray}
\frac{V^\prime}{V}&=&\frac{p}{\phi}, \label{c}  \\
\frac{V^{\prime\prime}}{V}&=&\frac{p(p-1)}{\phi^2}.
\end{eqnarray}
A small $p$ may be motivated from backreaction of the inflaton potential energy on heavy scalar fields \cite{Dong:2010in} as well as flux flattening in axion monodromy inflation \cite{Landete:2017amp}. 
In the following, we will assume $V/\Lambda \gg 1$ so that the brane effect is significant. We will see that this imposes an upper bound to $\Lambda$.
From Eqs.~(\ref{epsilon}) and (\ref{eta}), we have
\begin{eqnarray}
\epsilon &=& \frac{2p^2\Lambda}{a \phi^{p+2}} \label{ep}  \\
\eta &=& \frac{2 \Lambda p(p-1)}{a \phi^{p+2}}  \\
\frac{\eta}{\epsilon}&=&\frac{p-1}{p}. \label{et}
\end{eqnarray}
Inflation ends at $\phi = \phi_e$ when one of the slow-roll parameters becomes of order one, namely $\mbox{max}\{\epsilon,|\eta|\}= 1$. We can see from Eq.~(\ref{et}) that if $p=0.5$, $\epsilon = |\eta|$. However, if $p>0.5$, $\epsilon > |\eta|$. In this case, we have 
\begin{equation}
\phi^{p+2}_e=\frac{2p^2 \Lambda}{a}.
\end{equation}
On the other hand, If $0<p<0.5$, $\epsilon < |\eta|$, therefore
\begin{equation}
\phi^{p+2}_e=\frac{2 \Lambda p (1-p) }{a}.
\end{equation}
From Eq.~(\ref{efolds}), for $p>0.5$, we obtain
\begin{equation}
N=\int^{\phi_i}_{\phi_e}\frac{a \phi^{p+1}}{2p \Lambda}d\phi  = \frac{a \phi_i^{p+2}}{2p(p+2) \Lambda}-\frac{p}{p+2} .
\end{equation}
For $0<p<0.5$, we obtain
\begin{equation}
N=\int^{\phi_i}_{\phi_e}\frac{a \phi^{p+1}}{2p \Lambda}d\phi  = \frac{a \phi_i^{p+2}}{2p(p+2) \Lambda}-\frac{1-p}{p+2} .
\end{equation}
In both cases, comparing with $50 \lesssim N \lesssim 60$, we can neglect the contribution of $\phi_e$, therefore at horizon exit we have
\begin{equation}
\phi_i^{p+2}=\frac{2Np(p+2) \Lambda}{a}.
\label{phii}
\end{equation}
Substitute this result into Eq.~(\ref{ep}), we obtain
\begin{equation}
\epsilon = \frac{p}{N(p+2)}.
\label{epsilon2}
\end{equation}
From the above equation and Eqs.~(\ref{et}) and (\ref{index}), the spectral index is given by
\begin{equation}
n_s=1-\frac{2+4p}{N(p+2)}.
\label{indexf}
\end{equation}
Interestingly, this result does not depend on the parameters $a$ nor $\Lambda$ as long as our assumption $V/\Lambda \gg 1$ is valid. 
The spectral index as a function of $p$ from Eq.~(\ref{indexf}) is plotted in Fig.~\ref{fig7}. Depending on the value of $p$, the spectral index $n_s$ is more or less within the observational constraints  \cite{Akrami:2018odb}. 
The running spectral index is given by
\begin{equation}
\alpha = -\frac{2+4p}{N^2(p+2)},
\end{equation}
which is plotted in Fig.~\ref{running}. As we can see from the plot, the running spectral index is in accordance with the Planck data $|\alpha| \lesssim 0.01$ \cite{Akrami:2018odb}.
Subsituting Eq.~(\ref{phii}) into Eq.~(\ref{spectrum}), we obtain
\begin{equation}
P_R=\frac{1}{96 \pi^2}(2N)^{\frac{4p+2}{p+2}}p^{\frac{2p-2}{p+2}}(p+2)^{\frac{4p+2}{p+2}}\Lambda^{\frac{p-4}{p+2}}a^{\frac{6}{p+2}}=(5 \times 10^{-5})^2, 
\end{equation}
where CMB normalization is imposed. This gives a relation between $a$ and $\Lambda$ for some fixed $p$. Hence from Eq.~(\ref{phii}), we can obtain
\begin{equation}
\phi_i=8.66\times p^{\frac{1}{3}}[2N(p+2)]^{\frac{2}{3}}\Lambda^{\frac{1}{6}}.
\label{phibound}
\end{equation}
Since $V$ is decreasing during inflation, if our assumption $V/\Lambda \gg 1$ is satified at the end of inflation, say $V/\Lambda >100$, it is satisfied during inflation. For $p>0.5$, this imposes a condition
\begin{equation}
\Lambda^{\frac{1}{6}}<1.15 \times 10^{-2}\times 2^{\frac{p}{2p+4}}\left[ \frac{p}{2N(p+2)}\right]^{\frac{2p+1}{3p+6}}.
\end{equation}
This is plotted in Fig.~\ref{fig5}.
By using Eq.~(\ref{phibound}) and the upper bound of $\Lambda^{\frac{1}{6}}$, we can obtain an upper bound of $\phi_i$, which is plotted in Fig.~\ref{fig1}. We can see that for $0.5 \leq p \leq 2$ the first Swampland criterion is satisfied even for the upper bound of $\Lambda^{\frac{1}{6}}$. In order to check the second Swampland criterion, we plot the lower bound of $V^\prime/V$ at $\phi_i$ in Fig.~\ref{fig4}. Note that since $\phi$ decreases during inflation,  we can see from Eq.~(\ref{c}) that $V^\prime/V$ increases during inflation. This implies that if the second Swampland criterion is satisfied at $\phi_i$, it is satisfied during the whole period of inflation. 

\begin{figure}[t]
\centering
\includegraphics[width=0.7\columnwidth]{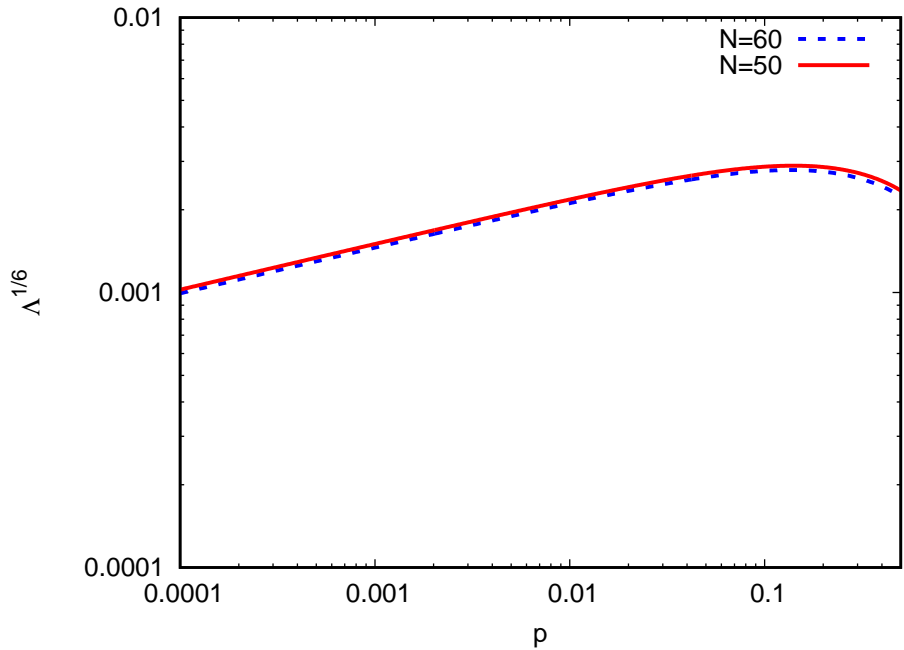}
 \caption{The upper bound of $\Lambda^{1/6}$ as a function of $p$ when $0.0001 \leq p \leq 0.5$ from the condition $V/\Lambda > 100$. Both number of e-folds $N=50$ and $60$ are shown.}
\label{fig6}
\end{figure}

\begin{figure}[t]
\centering
\includegraphics[width=0.7\columnwidth]{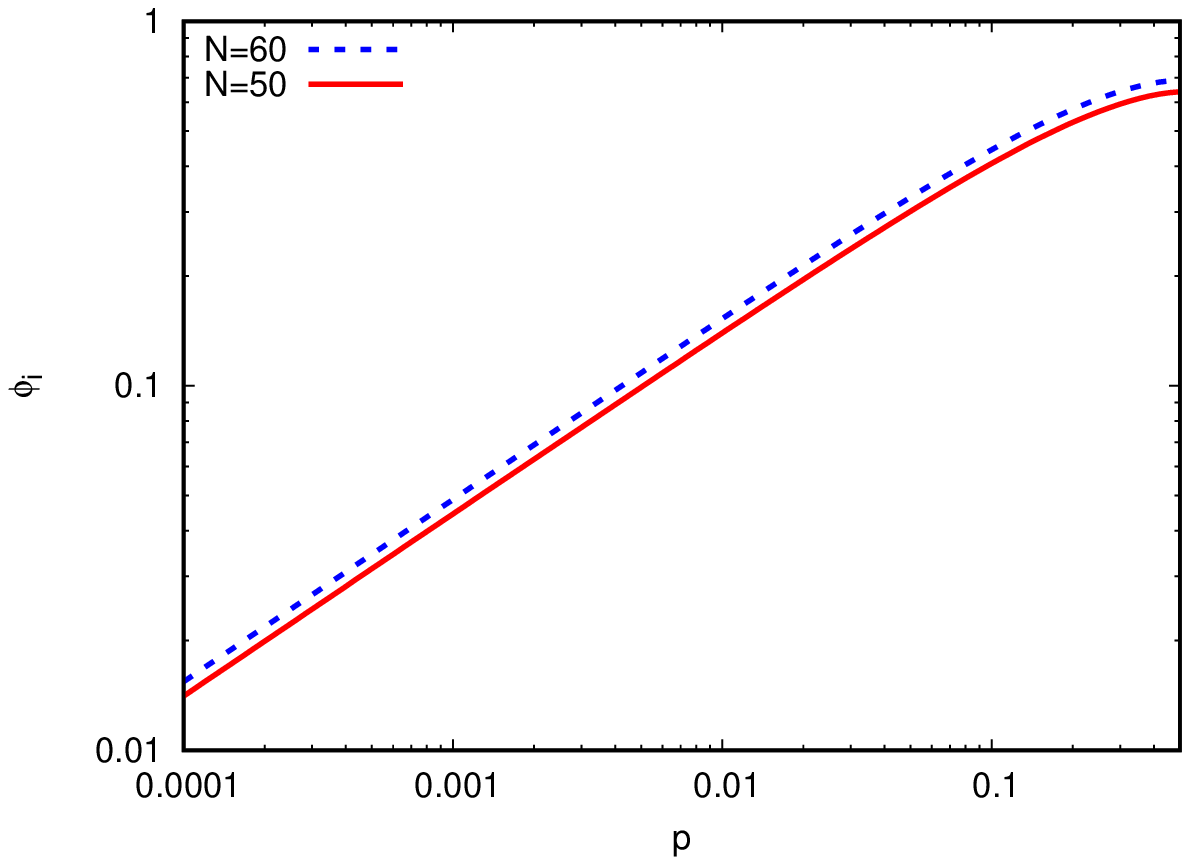}
 \caption{The upper bound of $\phi_i$ as a function of $p$ when $0.0001 \leq p \leq 0.5$ from the condition $V/\Lambda > 100$. Both number of e-folds $N=50$ and $60$ are shown.}
\label{fig2}
\end{figure}

\begin{figure}[t]
\centering
\includegraphics[width=0.7\columnwidth]{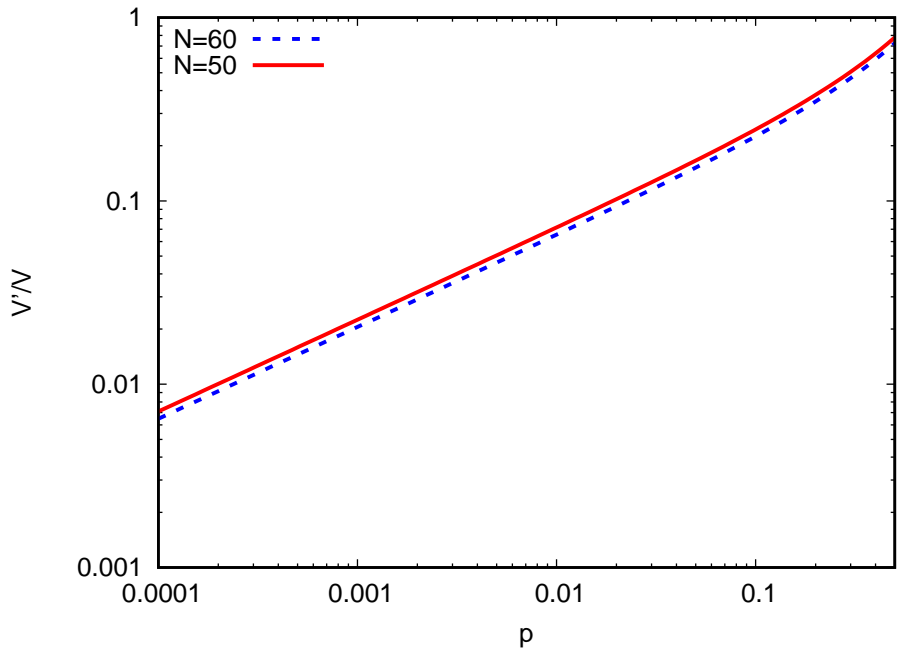}
 \caption{The lower bound of $V^\prime/V$ as a function of $p$ when $0.0001 \leq p \leq 0.5$ from the condition $V/\Lambda > 100$. Both number of e-folds $N=50$ and $60$ are shown. When we choose $\Lambda^{\frac{1}{6}}$ to be one order lower, $V^\prime/V$ is increased by one order. Since we can reduce $\Lambda^{\frac{1}{6}}$ by roughly six orders from its upper bound by Eq.~(\ref{reduce}), the Swampland de Sitter conjecture can be satisfied.}
\label{fig3}
\end{figure}

For $p<0.5$, the corresponding condition for $\Lambda^{\frac{1}{6}}$ is
\begin{equation}
\Lambda^{\frac{1}{6}}<1.15 \times 10^{-2}\times 2^{\frac{p}{2p+4}}\times p^{\frac{1}{6}}(1-p)^{\frac{p}{2p+4}}[ 2N(p+2)]^{-\frac{2p+1}{3p+6}}.
\end{equation}
This is plotted in Fig.~\ref{fig6} and the upper bound of $\phi_i$ is plotted in Fig.~\ref{fig2}. The lower bound of $V^\prime/V$ is plotted at $\phi_i$ in Fig.~\ref{fig3}. As we can see in Figs.~\ref{fig5} and \ref{fig6}, we have $\Lambda^{\frac{1}{6}} \lesssim 10^{-3}$ in the range $0.0001\leq p \leq 2$.  Combined with the lower bound of $\Lambda$ given in Section~\ref{sec2}, we obtain
\begin{equation}
1.92 \times 10^{-9}  \lesssim    \Lambda^{\frac{1}{6}} \lesssim 10^{-3}.
\label{reduce}
\end{equation} 
This means that $\Lambda^{\frac{1}{6}}$ can be many orders smaller than the upper bound.
Each time when we lower $\Lambda^{\frac{1}{6}}$ by one order from the upper bound, $\phi_i$ is reduced by one order, and $V^\prime/V$ is increased by one order.
Thus from Figs.~\ref{fig1}, \ref{fig4}, \ref{fig2}, and \ref{fig3} we can see that the Swampland criteria can be satisfied by judiciously choosing an appropriate $\Lambda^{\frac{1}{6}}$ within the allowed range.

\begin{figure}[t]
\centering
\includegraphics[width=0.7\columnwidth]{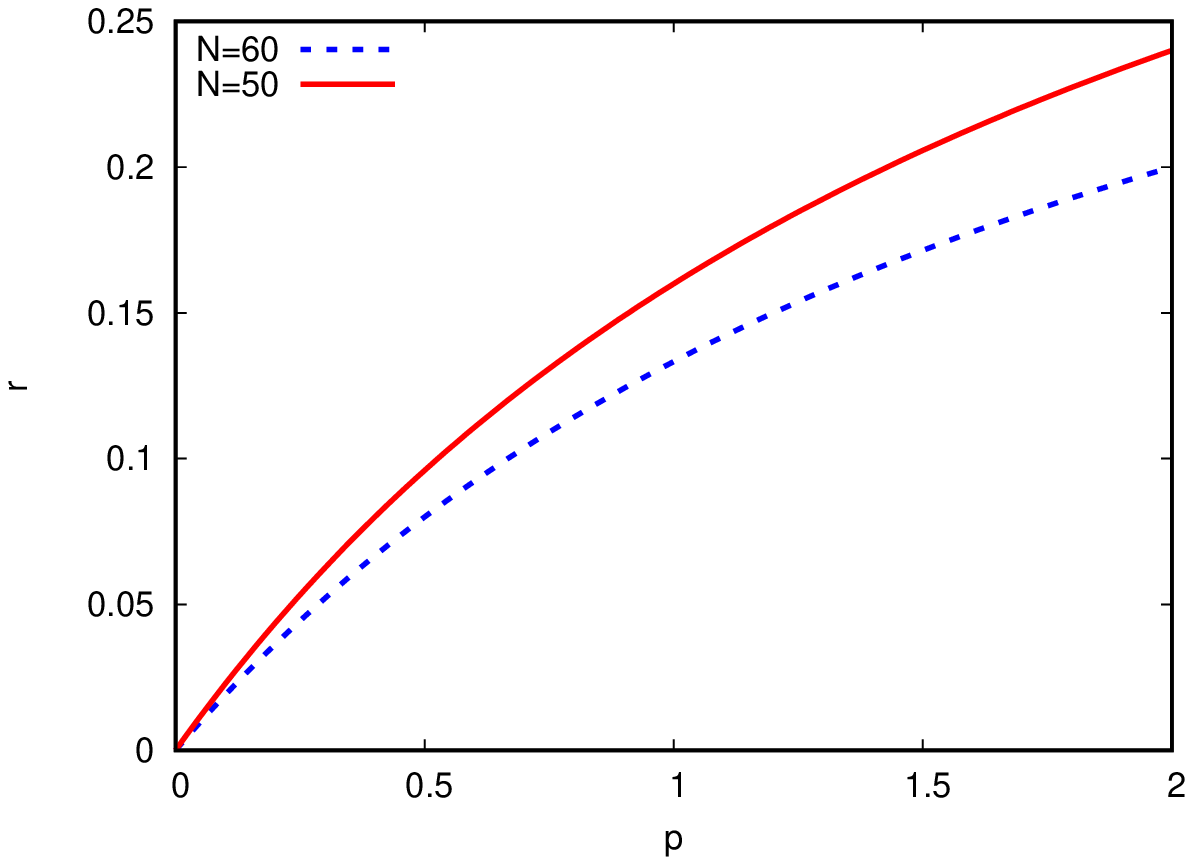}
 \caption{The tensor to scalar ratio $r$ as a function of $p$. Both number of e-folds $N=50$ and $60$ are shown.}
\label{fig8}
\end{figure}

\begin{figure}[t]
\centering
\includegraphics[width=0.7\columnwidth]{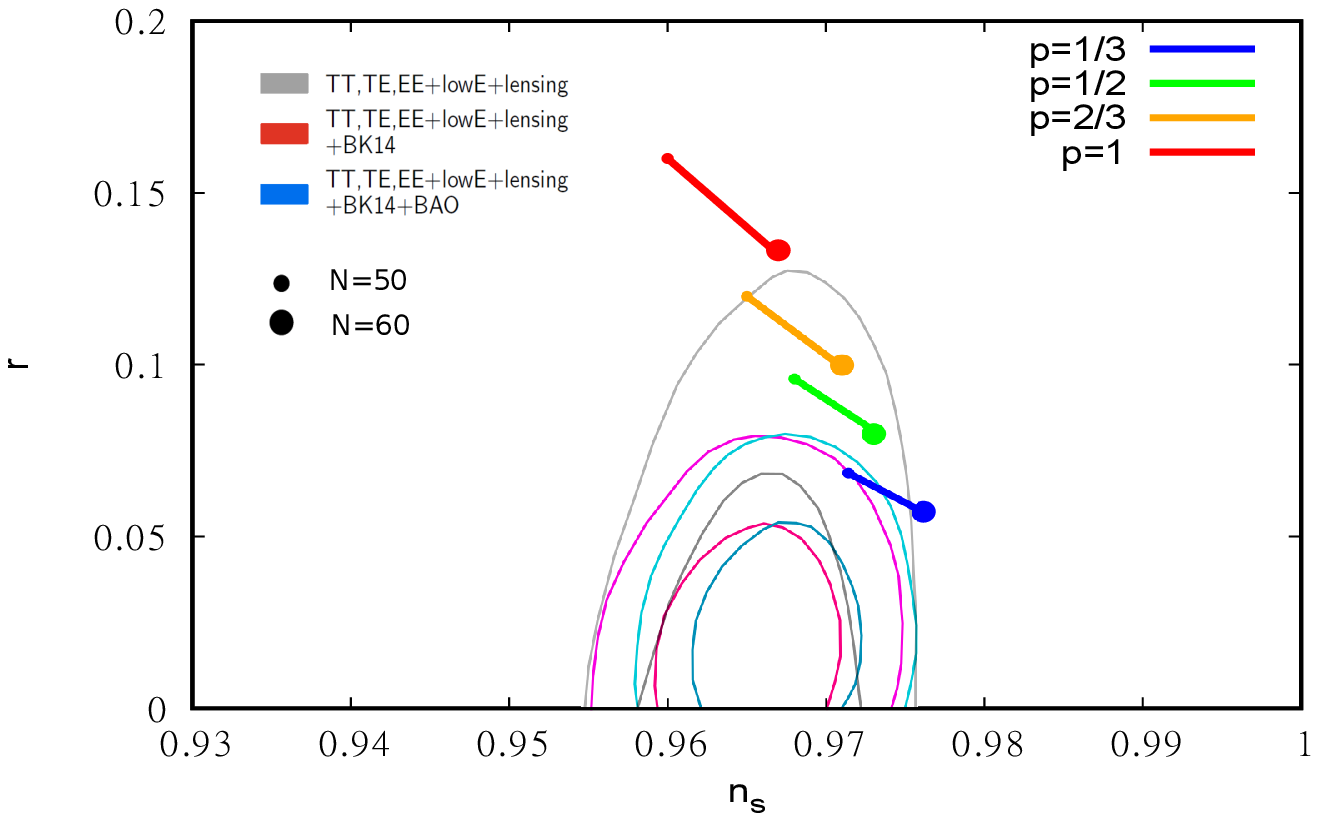}
 \caption{The spectral index $n_s$ and the tensor to scalar ratio $r$. The experimental constraints are from Ref.~\cite{Akrami:2018odb}.}
\label{fig100}
\end{figure}

The tensor to scalar ratio can be obtained from Eq.~(\ref{epsilon2}) as \cite{Langlois:2000ns,Bento:2008yx}\footnote{For conventional slow-roll inflation, or $V/\Lambda \ll 1$, the tensor to scalar ratio is given by $r=16 \epsilon$. Therefore for the same $\epsilon$, the tensor to scalar ratio $r$ would be bigger for $V/\Lambda \gg 1$.}
\begin{equation}
r=24 \epsilon = \frac{24 p}{N(p+2)},
\end{equation}
which is plotted in Fig.~\ref{fig8}. 
In order to compare the model with the latest experimental results, we plot the spectral index and the tensor to scalar ratio overlaid with the constraints taken from Ref.~\cite{Akrami:2018odb} in Fig.~\ref{fig100}. From the figure we can see that when we reduce $p \lesssim \mathcal{O}(1)$, the tensor to scalar ratio $r$ starts to enter a range within the experimental constraints with an acceptable value of the spectral index $n_s$.
\section{Conclusion and Discussion}
\label{con}
In this paper, we have calculated the field value, the spectral index, the running spectral index the tensor to scalar ratio for chaotic inflation on the brane with a potential of the form $V=a \phi^p$. In general, the Swampland criteria can be satisfied for a small $p$. The tensor to scalar ratio can be reduced to be within the experimental bounds via lowering the value of $p$ while the spectral index is within an acceptable value. Thus we conclude that single-field chaotic inflation on the brane where the primordial density perturbation is from inflaton fluctuations with the Bunch-Davies initial state is compatible with the Swampland criteria.
\section*{Acknowledgement}
This work is supported by the Ministry of Science and Technology (MOST) of Taiwan under grant numbers MOST 106-2112-M-167-001 (C. M. L.), MOST 107-2119-M-001-030 (K. W. N.), MOST-105-2112-M-007-028-MY3 (K. C.) and MOST-107-2112-M-007-029-MY3 (K. C.). We thank Dr. Yuta Hamada for a useful discussion on higher-dimensional versions of the conjecture.

\end{document}